\documentclass[prd,a4paper,twocolumn,showpacs,floatfix]{revtex4-1}

    \usepackage[utf8]{inputenc}
    \usepackage[T1]{fontenc}

    \usepackage{amssymb}
    \usepackage{amsmath}
    \usepackage{amsbsy}
    \usepackage{bm}
    \usepackage{graphicx}
 \usepackage{xcolor}

\begin{document}
\title{Photon enhancement in a homogeneous axion dark matter background}
\author{Ariel Arza}

\affiliation{Department of Physics, University of Florida, Gainesville, Florida 32611, USA}

\begin{abstract}
We study the propagation of photons in a homogeneous axion dark matter background. When the axion decay into two photons is stimulated, the photon field exhibits a parametric instability in a small bandwidth centered on one half of the axion mass. We estimate analytically the enhancement for both coherent and non-coherent axion fields and we find that this effect could be relevant in the context of miniclusters and galactic halos.
\end{abstract}	

\pacs{}

\maketitle

\section{Introduction}
Several ideas beyond the standard model of elementary particles have been performed to explain the dark matter of the universe at a fundamental scale. One of the most serious candidates is the QCD axion which was originally proposed in order to explain the conservation of the discrete symmetries $P$ and $CP$ in strong interactions  \cite{peccei}. In this theoretical framework, the axion is a pseudo-Goldston boson arising from the spontaneous breaking of the PQ (Peccei-Quinn) symmetry defined at some energy scale $f_a$. Below the QCD energy scale, the axion acquires its mass $m_a$ by nonperturbative effects. It is related to PQ symmetry breaking scale by
\begin{equation}
m_a\simeq6\times10^{-6}\text{eV}{10^{12}\text{GeV}\over f_a}. \label{axionmass}
\end{equation}
As the axion mass, all the couplings with the standard model are inversely proportional to $f_a$ and since the PQ symmetry breaking is supposed to be very high ($f_a\sim10^{12}\text{GeV}$), the axion is weakly coupled to the standard model particles. Besides this feeble interaction, one of the most important features of axion physics is that they could be efficiently  produced in the early universe by  non-thermal mechanisms \cite{wilczek2,abbott,dine}, such as the misalignment or from topological defect decays \cite{giddings,hindmarsh,chang}. Under this context, they can be abundant and cold enough to be an excellent dark matter candidate. On the other hand, pseudo-Nambu-Goldstone bosons emerging from the spontaneous breaking of global symmetries appear in several extensions of the Standard Model, and generically in all string compactifications \cite{witten}. They share many of the axion features, such as their small mass (also suppressed by the scale of the spontaneous symmetry breaking), their weak coupling to the standard model particles, and very importantly, they can also be produced in the early universe by the same non-thermal processes as axions so they are  cold dark matter candidates too \cite{masso,arias}. Due to all these similarities they are widely known as axion-like particles (ALPs), even though they do not feature an a priori relationship between the decay constant and their mass. One interesting attribute of axions (from now on we will refer as axions for both QCD axions and ALPs) from the experimental and observational point of view, is its coupling to two photons. The effective lagrangian of the axion-photon interaction is
\begin{equation}
{\cal L}={1\over2}\partial_\mu a\partial^\mu a-{1\over2}m_a^2a^2-{1\over4}F_{\mu\nu}F^{\mu\nu}+{1\over4}gaF_{\mu\nu}\tilde F^{\mu\nu}, \label{L1}
\end{equation}
where $a$ is the axion field and $g$ its coupling to two photons. $F_{\mu\nu}$ is the electromagnetic strength tensor $F_{\mu\nu}=\partial_\mu A_\mu-\partial_\nu A_\nu$, where $A_\mu$ is the electromagnetic field. $\tilde F^{\mu\nu}$ is the dual of the electromagnetic strength tensor and is defined as $\tilde F^{\mu\nu}={1\over2}\epsilon^{\mu\nu\rho\lambda}F_{\rho\lambda}$.

Axions have not been experimentally observed yet, so their existence is still in suspense, but because of their cosmological role as dark matter candidates, the experimental efforts to search for them in the corresponding parameter space have been intense. In the laboratory, axions are  searched for using the coupling to two photons described by Eq. (\ref{L1}). Especially, static magnetic fields have been strongly implemented \cite{sikivie,krauss,vanbibber}. Some running  experiments, such as haloscopes searches  \cite{ADMX}, already have access to the parameter space where they are cold dark matter candidates, and several new proposals plan to do so in the near future (ALPS-II \cite{ALPSII}, IAXO \cite{IAXO}, HAYSTAC \cite{HAYSTAC}, MADMAX \cite{MADMAX}, among others).

In this paper we are interested in studying the effects on an incident electromagnetic plane wave traveling within a homogeneous axion dark matter background.  We find an small window where the photon field experiences parametric resonance, we estimate the rate of the enhancement in coherent and non-coherent axion background and finally, we apply our results to astrophysical structures. Related topics were discussed recently in References \cite{yoshida}, but they do not consider non-coherent axion fields. Studies about the propagation of photons in an axion dark matter background had already been performed a few years ago in order to account for their effects in optical experiments \cite{espriu}, however the instability was not considered.

 The article is organized as follows:  in Section \ref{dos} we obtain amplitude equations for states excited by the process $a\rightarrow2\gamma$ and we analyze them numerically. Then, we solve the linear regime analytically finding the instability window and enhancement rate. In section \ref{tres} we estimate the results obtained in section \ref{dos} for coherent and non-coherent homogeneous axion fields. In section \ref{cuatro} we discuss the possible astrophysical relevance in axion miniclusters and galactic halos. 

\section{Electromagnetic plane wave in a homogeneous axion background} \label{dos}

\begin{figure}[t]
\raggedright
    \includegraphics[width=\linewidth]{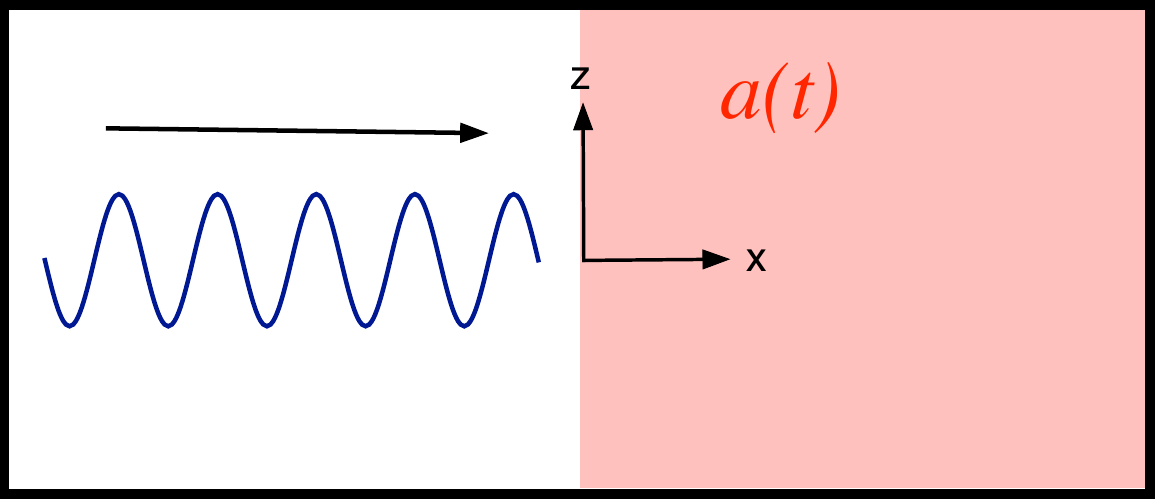}
\caption{Incident plane electromagnetic wave with wave vector $\vec k=\omega\hat x$ and linearly polarized in $\hat z$. The pink region corresponds to a homogeneous axion dark matter background}
\label{fig1}
\end{figure}

\begin{figure}[t]
\raggedright
    \includegraphics[width=\linewidth]{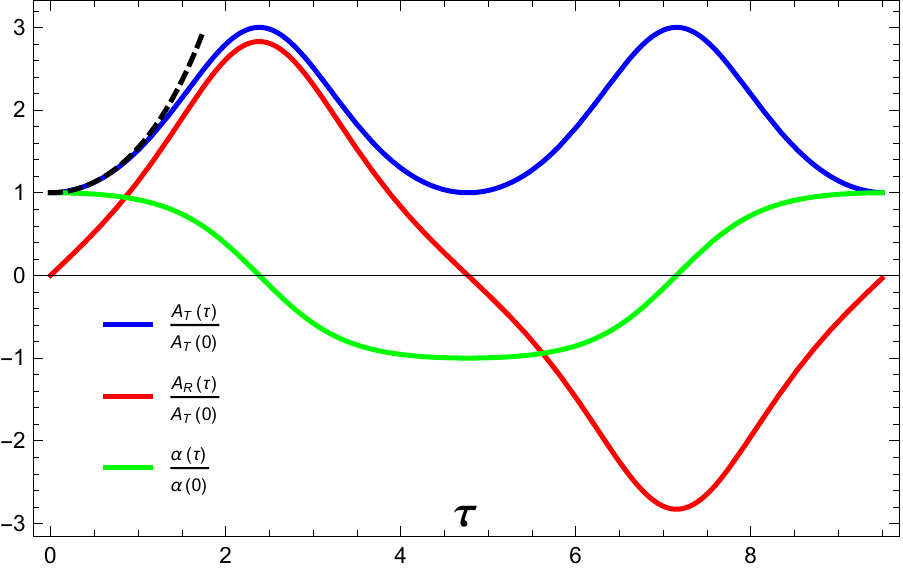}
\caption{Plot corresponding to the solution of Eqs. (\ref{eqAT1}), (\ref{eqAR1}) and (\ref{eqalpha1}) as a function of the dimensionless parameter $\tau=gm_a\alpha(0)t/2$ and initial conditions $A_T(0)=\alpha(0)/2$ and $A_R(0)=0$. The time dependent amplitudes $A_T$ and $A_R$ are normalized by $A_T(0)$ while $\alpha$ is normalized by $\alpha(0)$. The dashed black line corresponds to $A_T$ in the linear limit.}
\label{fig2}

\end{figure}

The equations of motion for photons in an homogeneous axion background can be written in the Coulomb gauge as
\begin{eqnarray}
\left(\partial_t^2-\nabla^2\right)\vec A &=& -g\partial_ta\vec\nabla\times\vec A, \label{eqA1}
\\
\left(\partial_t^2+m_a^2\right)a &=& g\partial_t\vec A\cdot\vec\nabla\times\vec A. \label{eqa1}
\end{eqnarray}
Let's consider an incident linearly polarized and plane electromagnetic wave with frequency $\omega$ as shown in Fig. \ref{fig1}. Inside the region filled with dark matter axions, the potential vector can be written as
\begin{equation}
\vec A(x,t)=\hat zA_T(t)e^{i\omega(x-t)}-i\hat yA_R(t)e^{-i\omega(x+t)}+c.c., \label{ansA}
\end{equation}
where the amplitudes $A_T(t)$ and $A_R(t)$ vary in time much slower than $e^{-i\omega t}$. On the other hand, the axion field has the form $a(t)=\alpha(t)e^{-im_at}+c.c.$, where the time dependent amplitude $\alpha(t)$ is introduced due to the feedback from the potential vector. The most interesting scenario is where the incident wave stimulates the process $a\rightarrow2\gamma$. In that case, the conservation of momentum and energy in the axion rest frame requires the condition $\omega=m_a/2$ to be satisfied. Eqs. (\ref{eqA1}) and (\ref{eqa1}) become
\begin{eqnarray}
\partial_tA_T &=& {1\over2}gm_a\alpha A_R^*, \label{eqAT1}
\\
\partial_tA_R^* &=& {1\over2}gm_a\alpha^* A_T, \label{eqAR1}
\\
\partial_t\alpha &=& -{1\over4}gm_aA_RA_T, \label{eqalpha1}
\end{eqnarray}
where we have neglected high frequency oscillation terms. The complete behavior of this system is shown in Fig. \ref{fig2}, where we have chosen the initial conditions $A_T(0)=\alpha(0)/2$ and $A_R(0)=0$ (the initial wave travels forward). We can see that the system is unstable increasing the electromagnetic field considerably. In order to compute analytically this instability, we focus in the linear behavior (see dashed black line in Fig. \ref{fig2}) which corresponds to the limit of a constant axion energy density. In such a case, the axion amplitude $\alpha(t)$ is considered as a slow oscillating function, which can be written as

\begin{equation}
\alpha(t)={\sqrt{\rho/2}\over m_a}e^{-i\epsilon t}, \label{axionamp1}
\end{equation}
where $\rho$ is the axion energy density and $\epsilon$ corresponds to small deviations of the axion frequency. Defining $A_T=a_Te^{-{i\over2}\epsilon t}$ and $A_R=a_Re^{-{i\over2}\epsilon t}$, Eqs. (\ref{eqAT1}) and (\ref{eqAR1}) leads to
\begin{equation}
\left[\partial_t^2-\left({g\sqrt{2\rho}\over4}\right)^2+{\epsilon^2\over4}\right]
\left[
\begin{array}{cc}
a_T
\\
a_R
\end{array}
\right] 
=0. \label{eqaTaR}
\end{equation}
We can see that for
\begin{equation}
-g\sqrt{\rho\over2}<\epsilon<g\sqrt{\rho\over2}, \label{instability}
\end{equation}
there is a parametric resonance where $a_T$ and $a_R$ grow exponentially in time being their complete solutions 
\begin{eqnarray}
a_T(t) &=& a_T(0)\left(\cosh(st)+{i\epsilon\over2s}\sinh(st)\right)  \label{aTsol1}
\\
a_R(t) &=& {\sigma\over s}a_T(0)^*\sinh(st),  \label{aRsol1}
\end{eqnarray}
where
\begin{equation}
\sigma={g\over2}\sqrt{\rho\over2} \ \ \ \ \text{and} \ \ \ \ s=\sqrt{\sigma^2-\epsilon^2/4}. \label{sigma1}
\end{equation}
Inside the instability window (\ref{instability}), the potential vector grows exponentially in time according to Eqs. (\ref{aTsol1}) and (\ref{aRsol1}), and the electromagnetic signals (e.g. the power) will be enhanced at the rate $e^{\gamma t}$, where
\begin{equation}
\gamma\sim2\sigma=g\sqrt{\rho\over2}. \label{gamma1}
\end{equation} 
The signal is a spectral line centered at $m_a/2$ with a bandwidth of $\Delta\omega=g\sqrt{\rho/2}$ (because of the parametric resonance window (\ref{instability})) and amplitude 
\begin{equation}
\left.{dP\over d\omega}\right|_{\omega=m_a/2}\approx{e^{\gamma t}\over4}\left.{dP_0\over d\omega}\right|_{\omega=m_a/2}, \label{eq4}
\end{equation}
where $dP_0/d\omega$ is the power spectrum of the incident beam.

To finish this section, we want to make some comments about the validity of our results in the linear approximation. To estimate the moment in which the exponential behavior is no longer valid, we apply $\partial_t$ to Eq. (\ref{eqAT1}). Using Eqs. (\ref{eqAR1}) and (\ref{eqalpha1}) we find
\begin{equation}
\partial_t^2A_T={1\over4}g^2m_a^2(|\alpha|^2-{1\over2}|A_R|^2)A_T. \label{eq1}
\end{equation}
We can see that the instability works until a time in which 
\begin{equation}
|A_R|^2<2|\alpha|^2. \label{eq2}
\end{equation}
After that, the right-hand side of (\ref{eq1}) becomes negative and $A_T$ starts to oscillate. In the linear regime, after a sufficient time $t$, we can approximate $|A_R|^2\approx {1\over4}|A_T(0)|^2e^{\gamma t}$ and $|\alpha|^2\approx|\alpha(0)|^2$. Condition (\ref{eq2}) translates to
\begin{equation}
e^{\gamma t}<8{|\alpha(0)|^2\over|A_T(0)|^2}. \label{eq3}
\end{equation}
Since the maximum enhancement is closely related with the validity of the linear regime, it must be proportional to ${\cal A}=|\alpha(0)|^2/|A_T(0)|^2$. In Fig. \ref{fig2} we use ${\cal A}=4$, however it could be much bigger considering that $|\alpha(0)|^2$ is given by the dark matter energy density.

\section{Enhancement in a coherent and non-coherent axion background} \label{tres}
Consider a dispersive axion background with frequency distribution $F(\omega_a)$ and total energy density $\rho_T$. This distribution is normalized as 
\begin{equation}
\int F(\omega_a)d \omega_a=1 \label{norm}
\end{equation}
and characterized by a bandwidth $\Delta\omega_a$ given by
\begin{equation}
\Delta\omega_a=m_a\delta v^2, \label{Deltaomega1}
\end{equation}
where $\delta v$ is the velocity dispersion. On the other hand, we have seen in the previous section that Eq. (\ref{instability}) defines the bandwidth
\begin{equation}
\delta\omega_a=g\sqrt{2\rho} \label{deltaomega1}
\end{equation}
in the axion frequency where axions produce parametric resonance. When $\delta\omega_a\sim\Delta\omega_a$ or greater, the axion field is coherent enough to matching $\rho=\rho_T$. In this case, Eq. (\ref{gamma1}) becomes
\begin{equation}
\gamma^{(\text{c})}\sim g\sqrt{\rho_T\over2}. \label{gammac}
\end{equation}
However, when $\delta\omega_a\ll\Delta\omega_a$, the axion field is considered as non-coherent for our purposes, and only a small axion population will be engaged with the instability. The effective energy density of this population is given by
\begin{equation}
\rho=\rho_TF(\omega_a)\delta\omega_a. \label{rho1}
\end{equation}
Combining Eqs. (\ref{deltaomega1}) and (\ref{rho1}), we find that the effective density is
\begin{equation}
\rho=2g^2\rho_T^2F(\omega_a)^2. \label{rho2}
\end{equation}
Replacing (\ref{rho2}) into (\ref{sigma1}), and assuming that $F(\omega_a)\sim\Delta\omega_a^{-1}$, we have
\begin{equation}
\gamma^{(\text{nc})}\sim {g^2\rho_T\over\Delta\omega_a}. \label{gamma2}
\end{equation}

\section{Possible astrophysical relevance} \label{cuatro}

\begin{figure}[t]
\raggedright
    \includegraphics[width=\linewidth]{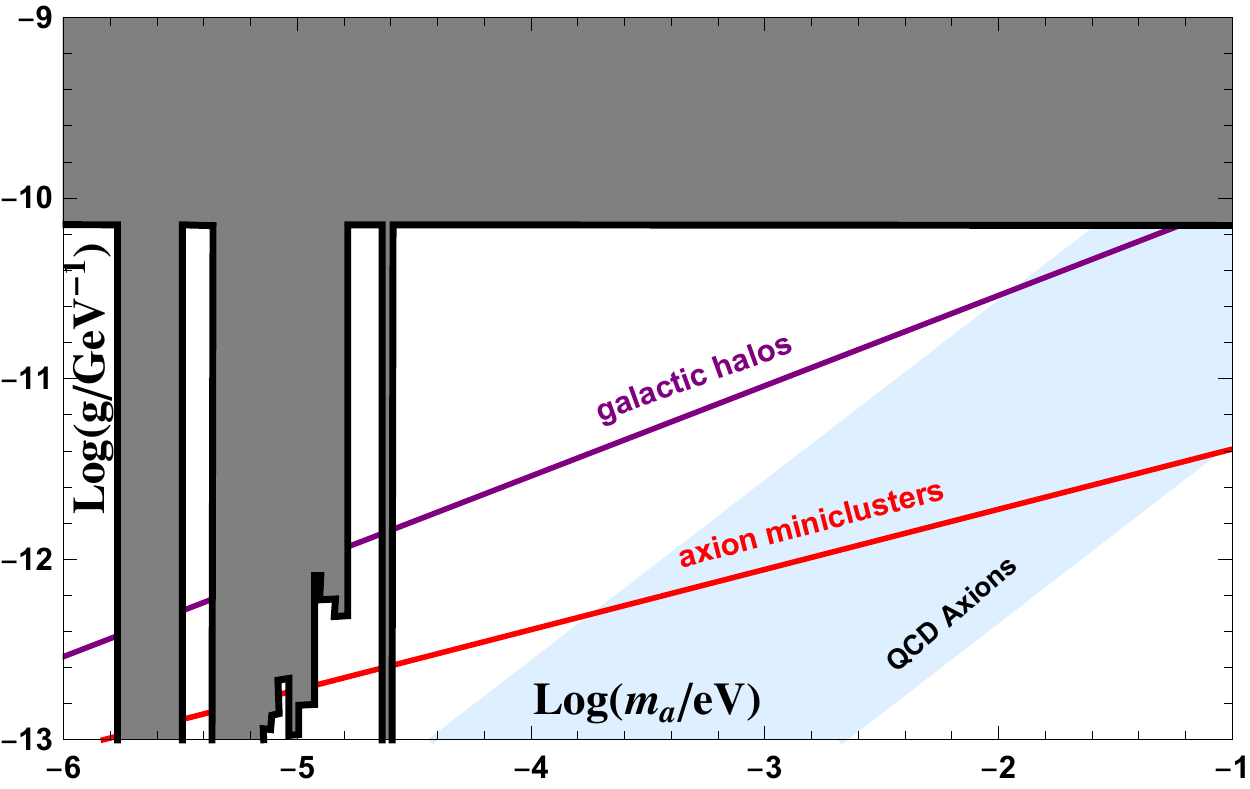}
\caption{Picture of the axion search in a small region of the parameter space $(m_a,g)$. The gray region is the space of parameters excluded by experiments (see Ref. \cite{irastorza}). Condition (\ref{criterion1}) is satisfied above the red and purple line for axion miniclusters and galactic halos assuming that the caustic ring model is correct. These lines were plotted using Eqs. (\ref{gmc}) and (\ref{gh}).}
\label{fig3}
\end{figure}

In order to figure out the parameter space $(m_a,g)$ where this effect could be important, we focus in two different astrophysical scenarios characterized by a large dark matter energy density; axion miniclusters and galactic halos. As a criterion, we establish that all the dark matter is composed by axions (or ALPs) and that parametric resonance plays a significative role when 
\begin{equation}
\gamma d>1, \label{criterion1}
\end{equation}
where $d$ corresponds to the distance traveled by the incident photon within the axion dark matter background, which is actually the maximum time during which the parametric resonance can occur. Using some typical physical parameters, we plot in Fig. \ref{fig3} the regions in the parameter space where condition (\ref{criterion1}) is satisfied for both axion miniclusters and galactic halos. More details of these results will be explained in the following.

\subsection{Axion Miniclusters}

Axion miniclusters have been well motivated if inflation occurs before PQ phase transition \cite{hogan}. In this case, inhomogeneities of the axion fluid form bound objects by gravitational instabilities. The typical total energy densities of the axion miniclusters is about $\rho_{mc}\sim10^{-18}\text{gr}/\text{cm}^3$ \cite{chang}. For axion miniclusters we assume that the axion background is coherent enough, for instance it could be considered as a Bose Einstein condensate \cite{sikivieBEC,hertzberg2}. In this case the exponential rate is just $\gamma\sim g\sqrt{\rho_{mc}/2}$ (see Eq. (\ref{gammac})). The typical size is
\begin{equation}
d_{mc}\sim t_1{R(t_{eq})\over R(t_1)}, \label{sizemc1}
\end{equation} 
where $R(t)$ is the scale factor and $t_{eq}\sim2\times10^{12}\text{s}$ is the cosmological time of matter-radiation equality. $t_1$ corresponds to the cosmological time when the axion field starts to oscillate and is defined by $m_a(t_1)t_1\sim1$. In the most general scenario (see \cite{arias}), the axion energy density today can be expressed as
\begin{equation}
\rho_{a,0}\simeq0.58\rho_\text{CDM}{m_a\over 10^{-5}\text{eV}}\left(t_1\over10^{7}\text{s}\right)^{1/2}\left(10^{-11}\text{GeV}^{-1}\over g\right)^2, \label{rhotoday}
\end{equation}
where $\rho_\text{CDM}$ is the today dark matter energy density. From Eq. (\ref{rhotoday}) we can find $t_1$ assuming that all the dark matter are axions, we have
\begin{equation}
t_1\sim3\times10^7\text{s}\left(10^{-5}\text{eV}\over m_a\right)^2\left(g\over10^{-11}\text{GeV}^{-1}\right)^4. \label{t1}
\end{equation}
Replacing Eq. (\ref{t1}) into Eq. (\ref{sizemc1}) and using $R(t)\sim t^{1/2}$, we have
\begin{equation}
d_{mc}\sim2\times10^{20}\text{cm}\left(10^{-5}\text{eV}\over m_a\right)\left(g\over10^{-11}\text{GeV}^{-1}\right)^2. \label{dmcALPs}
\end{equation}
Condition (\ref{criterion1}) gives us
\begin{equation}
g>1.89\times10^{-13}\text{GeV}^{-1}\left(m_a\over10^{-5}\text{eV}\right)^{1/3}. \label{gmc}
\end{equation}

\subsection{Galactic Halos}

The caustic ring halo model \cite{duffy} could be very interesting in this context because predicts regions in galaxies with large energy densities and very small velocity dispersion. For instance, it is estimated that the 5th caustic ring in the Milky Way, which is located very close to the sun, features an energy density of 1.5$\times 10^{-24}$gr/cm$^3$ and a velocity dispersion about 53m/s or less in a extension of about 130pc \cite{sikivie2}. In addition, there is no reason to think that other caustic rings in galactic halos can not have larger densities and widths. Considering reasonable parameters such as an energy density $\rho_h\sim1\text{GeV}/\text{cm}^3$, a velocity dispersion $\delta v_h\sim10^{-8}c$ (where $c$ is the speed of light) and a width $d_h\sim1\text{kpc}$, condition (\ref{criterion1}) becomes
\begin{equation}
g>9.13\times10^{-13}\text{GeV}^{-1}\left(m_a\over10^{-5}\text{eV}\right)^{1/2}. \label{gh}
\end{equation}
This expression was found using Eq. (\ref{gamma2}) which is more appropriate in this context where the axion field is non-coherent. 

As a final remark, It is very important to make clear that in this manuscript we are only considering homogeneous axion backgrounds. Inhomogeneities  could change partially or drastically our results.

\section*{Acknowledgments}

We would like to thank Pierre Sikivie for important discussions and Paola Arias for comments. This work was supported by the Chilean Commission on Research, Science and Technology (CONICYT) under grant 78180100 (Becas Chile, Postdoctorado).

\end{document}